\documentclass[10pt,conference]{style_file}
\usepackage{color}
\usepackage{url}
\usepackage{graphicx} 
\usepackage{epstopdf}
\usepackage{amsfonts}
\usepackage{graphicx}
\usepackage{amsmath}
\usepackage{times}
\usepackage{latexsym}
\usepackage{subfigure}
\usepackage{bm}
\usepackage{amssymb}
\usepackage{array}
\usepackage{setspace}
\usepackage{mathrsfs}
\usepackage{epsfig}
\usepackage{multicol}
\usepackage{subfigure}
\usepackage{mathtools}
\usepackage{float}
\usepackage{multirow}
\usepackage[utf8]{inputenc} 

\IEEEtriggeratref{26} 

\graphicspath{{./figures/}}

\usepackage[noadjust]{cite} 
\usepackage{graphicx}

\begin{document}

\title{5G MIMO Data for Machine Learning:\\Application to Beam-Selection using Deep Learning}


\author{\authorblockN{Aldebaro Klautau, Pedro Batista, }
\authorblockA{Dep. of Comp. and Telecomm. Eng.\\
Federal University of Par\'a\\
Belem, PA, 66075-110, Brazil \\
Emails: \{aldebaro,pedro\}@ufpa.br}
\and 
\authorblockN{Nuria Gonz\'alez-Prelcic,}
\authorblockA{Dep. of Signal Theory and Comm.\\
University of Vigo\\
Vigo, 36310, Spain\\
Email: nuria@gts.uvigo.es}
\and 
\authorblockN{Yuyang Wang and Robert W. Heath Jr.}
\authorblockA{Dep. of Elec. and Computer Eng.\\
The University of Texas, Austin\\
Austin, TX, 78712-1084, USA\\
Emails: \{yuywang,rheath\}@utexas.edu}
}
%

\maketitle

\begin{abstract}
The increasing complexity of configuring cellular networks suggests that machine learning (ML) can effectively improve 5G technologies. Deep learning has proven successful in ML tasks such as speech processing and computational vision, with a performance that scales with the amount of available data. The lack of large datasets inhibits the flourish of deep learning applications in wireless communications. This paper presents a methodology that combines a vehicle traffic simulator with a ray-tracing simulator, to generate channel realizations representing 5G scenarios with mobility of both transceivers and objects. The paper then describes a specific dataset for investigating beam-selection techniques on vehicle-to-infrastructure using millimeter waves. Experiments using deep learning in classification, regression and reinforcement learning problems illustrate the use of datasets generated with the proposed methodology.
\end{abstract}

\section{Introduction}

Machine learning (ML) has been applied to a large variety of problems in telecommunications, which include network management, self-organization, self-healing and physical layer (PHY) optimizations~\cite{jiang_machine_2017,klaine_survey_2017}. 
Deep learning (DL), a special category of ML, has been responsible for several recent performance breakthroughs in areas such as speech processing and computational vision~\cite{Lecun2015}. The success in other domains motivates the application of DL to communication problems~\cite{oshea_introduction_2017,he_integrated_2017,han_two-dimensional_2017,west_deep_2017,atallah_deep_2017,atallah_reinforcement_2017,xu_deep_2017,ferreira_multi-objective_2017,Ekman2017}.
While DL can be applied to any ML problem, its niche has been applications with large amount of data. 
The reason is that 
DL scales well with the amount of data and model complexity~\cite{Lecun2015}. 

In many DL application domains, the data is abundant or has a relative low cost.
For example, the DL-based text-to-speech system presented in \cite{Shen2018}, which represents the
state-of-the-art, achieves quality close to natural
human speech after being trained with a reasonable amount of digitized speech.
 In contrast, the research and development of 5G lower layers has to deal with a relatively limited amount of data. 
For example, mmWave measurements
for 5G MIMO research demands very expensive equipment and, eventually, elaborate outdoor measurement campaigns~\cite{maccartney_flexible_2017}. 
The lack of freely available data impairs the data-driven lines of investigation.

This paper presents a methodology for channel data generation in 5G millimeter wave (mmWave) multiple-input multiple-output (MIMO) scenarios~\cite{heath_overview_2016}.
The goal is to facilitate the investigation of ML-based problems related to the PHY of mmWave MIMO in 5G.
The presented methodology simplifies creating data in complicated (and potentially realistic) mobility scenarios through repeatedly invoking a traffic simulator and a ray-tracing simulator.
In the current context, generating propagation channel data is a reasonable way to alleviate the data scarcity
while benefiting from the accuracy associated to ray-tracing (RT)~\cite{Rappaport14,fuschini_ray_2015}. For instance, RT can cope with 5G requirements such as \emph{spatial consistency}, which has
been a challenge to traditional stochastic modeling~\cite{rumney_critical_2017,turkka_spatial_2016}. 
The simulated datasets do not substitute but complement data from measurements, which can 
improve and validate simulated data and statistical channel models, as they become available.
The paper also presents concrete examples of usage for the generated datasets via  experiments with DL for beam-selection in vehicle-to-infrastructure (V2I) mmWave communications.
Given the current amount of data is  limited, it is out of the scope of this paper 
to investigate performance of specific DL architectures. The goal is to illustrate instead the flexibility
provided by the data generation methodology.
This methodology can be used in applications other than V2I, as well
as to create datasets for ML problems concerning classification, regression, clustering and
time-based \emph{sequence} recognition~\cite{geron2017hands}.

Several ML techniques have been applied to the PHY processing (see, e.\,g.,~\cite{Wang17} and references therein). The large majority of previous work rely on simulations. For example, 
simulations are used in \cite{OShea17}, which present 
unsupervised DL architectures based on autoencoders for MIMO schemes. 
In~\cite{Mismar2018}, a classifier based on boosted trees 
was applied to the optimization of handovers between sub-6~GHz and mmWave radios using simulated channels.
In such cases, in which measurements are not available, our methodology can provide reasonably accurate channel data and advanced tools for modeling mobility. The generated datasets are especially useful when spatial consistency and time evolution are important to assess the ML technique.


The rest of the paper is organized as follows. The methodology for data generation is presented in Section~\ref{sec:data_generation}.
Beam-selection is the topic of
Section~\ref{sec:beam_selection}, in which a brief literature review is included. We discuss experiments
illustrating deep learning using the V2I dataset in Section~\ref{sec:experiments}, which is followed by the conclusions in 
Section~\ref{sec:conclusion}.

%
%
%
%

\section{Methodology for Data Generation}
\label{sec:data_generation}

In this section we describe the proposed methodology for data generation,
which has to take in account the challenges of mmWave channel modeling
such as the large signal bandwidths and prominent impact of scattering~\cite{Rappaport14}.
Other issues raise due to the intended application in 5G MIMO.
In scenarios of high mobility such as V2I, the channel evolution over time is important
to assess, for example, beam tracking techniques~\cite{va_impact_2017}.
For instance, 
mmWave communication when the vehicle speed is 35~m/s may have to cope
with a signal fading rate of 44~dB/s~\cite{maccartney_flexible_2017}.
In fact, difficult channel modeling in complex scenarios is the
first issue highlighted in~\cite{Wang17} when discussing challenges related to DL in wireless
communications. 
The following paragraphs discuss how RT and traffic simulators are used
in our methodology for circumventing issues when
generating ML datasets that depend on mmWave MIMO modeling.

\subsection{Ray-tracing simulation for mmWave MIMO channels}\label{sec:raytracing} 

\begin{figure}[htb]
\centering
\includegraphics[width=\columnwidth]{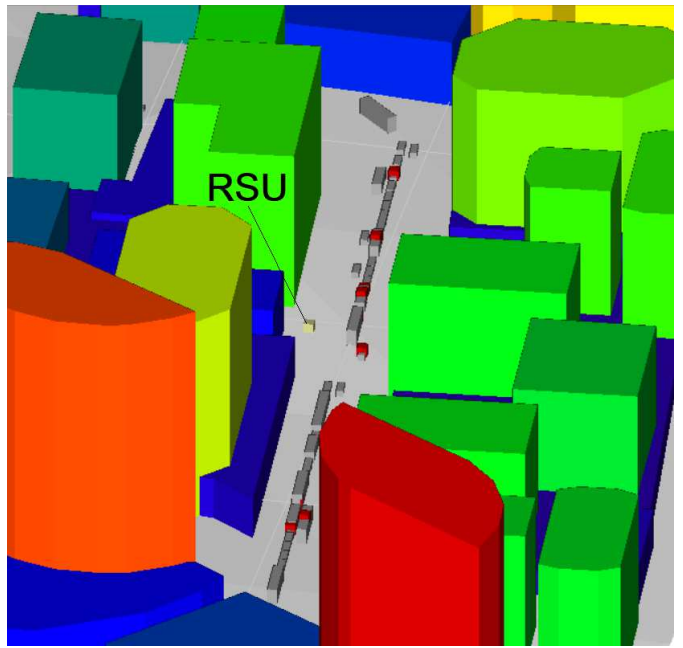}
\caption{Urban canyon scenario in a 3-d ray-tracing simulator with 
vehicles of distinct sizes randomly positioned. The building color indicates height and corresponds to a range from 0 (blue) to 101 meters (red).\label{fig:vehicles}}
\end{figure}

RT is considered a promising simulation strategy for 5G (see, e.\,g.~\cite{fuschini_ray_2015}).
RT can provide very accurate results~\cite{Rappaport14,stabler_mimo_2009_ak}
but its computational cost increases exponentially with the maximum
allowed number of reflections and diffractions~\cite{arikawa_simplified_2014}. Another issue of RT is that the generated channels are \emph{site specific}, depending on the specific propagation environment.
%
%
Besides, for improved RT accuracy, the scenario should be reasonably detailed. For example, outdoor scenarios 
require the detailed specification (including size, geometry, material, etc.) of buildings, vehicles, people and objects of interest such as a roadside unit (RSU) for V2I, as illustrated in Fig.~\ref{fig:vehicles}.
The geometrical aspects of the environment must be informed together with the
corresponding electromagnetic parameters such as the \emph{scattering coefficient} ($S$) for each material~\cite{fuschini_ray_2015}. 
Given the simulation scenario, a RT simulator projects rays in the three-dimensional angular space with a predefined spacing. Then, the paths are ranked according to the received power of each ray.
A detailed enough scenario description is 
the first challenge for RT usage. Another one is an appropriate modeling of the propagation channel,
which has to take in account, for example, the scattering of mmWave signals.


\begin{figure}[htb]
\centering
\includegraphics[width=\columnwidth]{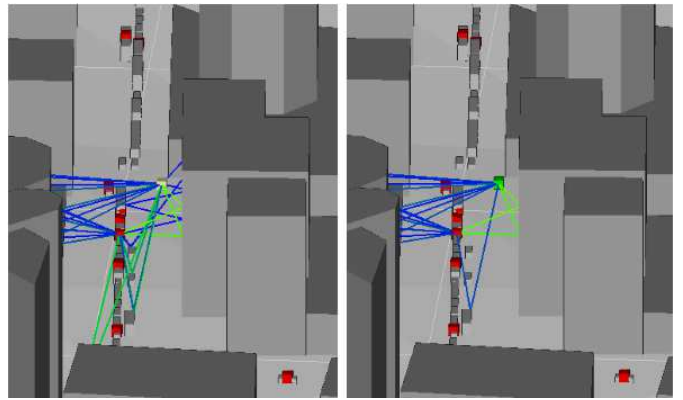}
\caption{Rays obtained in a traffic jam situation. 
The left figure shows all 25 most significant rays reaching a vehicle while the right one shows the subset (8 rays) corresponding to the ``diffuse scattered'' rays. The zoomed figure indicates there are three clusters of diffused rays reaching this receiver.\label{fig:diffusion_rays}} 
\end{figure}

\emph{Diffuse scattering} (DS) is an important feature of mmWave channel simulators~\cite{fuschini_ray_2015,solomitckii_characterizing_2016,insite}. This feature
can enrich the channel realizations and
minimize the chances of bias due to a limited number of specular rays, as found when materials are smooth.
The computational cost increases though with parameters such as the maximum allowed number of DS reflections ($N_{\textrm{max}}^{\textrm{DS}}$).
%
%
To illustrate DS, Fig.~\ref{fig:diffusion_rays} shows an example\footnote{For visualization, the camera in Fig.~\ref{fig:diffusion_rays} is rotated with respect to Fig.~\ref{fig:vehicles}.} of rays obtained in a simulation with a traffic jam (vehicles with receivers are marked in red) using 60~GHz. In this case, the simulation time increased by a factor of three when enabling DS with $N_{\textrm{max}}^{\textrm{DS}}= 2$.

A RT simulator may  support mobility, for example, allowing the receiver to follow a trajectory with
a given speed. Supporting the changing position of scatterers and blockers, though, complicates 
the required RT optimizations. Therefore, it is more common for a RT simulator to allow receivers to move, but not objects that can influence the rays. 
Simulating mobility then requires 
repeatedly invoking the simulator with the specification of a \emph{scene}. 
This is the case of Remcom's Wireless InSite~\cite{insite}, which is the RT simulator used in this paper.

\subsection{Spatial consistency and time evolution requirements}

While traditional \emph{drop-based} stochastic models have been extremely useful in the design of communication systems,
their application in the context of 5G  has been criticized with respect to spatial consistency~\cite{rumney_critical_2017,metis2015}.
For example, the results presented in~\cite{turkka_spatial_2016} indicate that
the 3GPP three-dimensional (3-d) geometry-based stochastic channel model underestimates the performance
of massive MIMO systems in line-of-sight (LOS) scenarios, while overestimating the performance
 of multi-user MIMO in specific ultra-dense scenarios with non-LOS (NLOS).
%
%

State-of-the-art stochastic and hybrid models for 5G have been incorporating features that aim at improved spatial consistency~\cite{molisch_spatially_2016,wang_extension_2016}.
Examples are the modeling techniques in NYUSIM~\cite{nyusim} and QuaDRiga~\cite{Jaeckel14}, as well the three techniques detailed in~\cite{3gpp-R1-161622}.
A classical alternative to stochastic models is RT~\cite{mckown_ray_1991,fuschini_ray_2015}, which is the simulation technique adopted
in this paper. RT is able to generate data with two key requirements for ML datasets with simulated channel realizations: spatial consistency and history of time evolution. 

In our methodology, the outputs of the simulators are periodically stored as ``snapshots'' (or \emph{scenes}) over time $t=n T_{\textrm{sam}}$, where $T_{\textrm{sam}}$ is the sampling period and $n \in \mathbb{Z}$. 
A \emph{scene} ${\cal{S}}(t)$ can contain multiple transmitters and receivers, which facilitates using the
datasets to investigate multiuser and other MIMO problems. In post-processing stages (an example is provided in Section~\ref{sec:postprocessing}), the information in ${\cal{S}}(t)$ is used
to model, for example, MIMO channels $\mathbf{H}(t)$.
A scene ${\cal{S}}(t)$ can potentially contain all \emph{out-of-band} information that the user gathered, including the ones provided by the RT and traffic simulators, such as position, vehicle dimension, angles of arrival, gains, etc. Section~\ref{sec:experiments} will discuss concrete examples on how the information in ${\cal{S}}(t)$ can be used
in beamforming applications.

For improved scene diversity and given the relatively high computational cost of a RT simulation,
we extract observation windows (or \emph{episodes}) at distinct instants. Specifically, instead of always consecutively extracting a scene along the whole simulation, episodes of duration $T_{\textrm{epi}}$ are obtained, each with $N_{\textrm{sce}}=\lfloor T_{\textrm{epi}} / T_{\textrm{sam}} \rfloor$ scenes. For example, an episode starting at time $t_0$ will be composed by a sequence of scenes $\{ {\cal{S}}(t), t=t_0, t_0+
T_{\textrm{sam}}, \ldots, t_0+(N_{\textrm{sce}}-1)T_{\textrm{sam}} \}$. 
For facilitating parallel processing, a dataset with $N_{\textrm{epi}}$ episodes can be organized as a TensorFlow TFRecord~\cite{geron2017hands}.

Keeping the channel variation over time enables investigating 
algorithms that take in account the channel dynamics. For more realistic simulations,
the mobility can be controlled by a specialized software as described in the next subsection.

\subsection{Integration of traffic and ray-tracing simulators}


Vehicle and pedestrian \emph{traffic simulators} provide flexibility to
investigate the impact of mobility in V2I and related applications. We describe the integration between the open source \emph{Simulation of Urban MObility} (SUMO) traffic simulator~\cite{SUMO2012} and Wireless InSite. There is extensive support to the use of SUMO with
network simulators such as OMNeT++, but the novel integration with RT facilitates studies targeting the mmWave PHY. 

The main role of the traffic simulator is to facilitate modeling mobility, especially the motions of both transceivers and potential scatters in the environment. It is possible to
directly get the necessary data only with the RT simulator but this may require considerable
effort if the scenario is complicated. Traffic simulators are specialized tools with plenty of features to describe vehicles with distinct characteristics, interaction with pedestrians, etc.
Adopting the right tool to model mobility enables the user to depart from simplistic scenarios, 
such as those in which all vehicles have constant speed.
Using a specialized traffic simulation tool to decouple the mobility specification from the RT simulation, simplifies the experiment configuration and grants to the user flexibility, for instance, to impose trajectories to any object or person,
use distinct speeds, etc. Also, the orientation of objects such as antennas can be automatically adjusted. 

\begin{figure}[htb]
\centering
\includegraphics[width=\columnwidth]{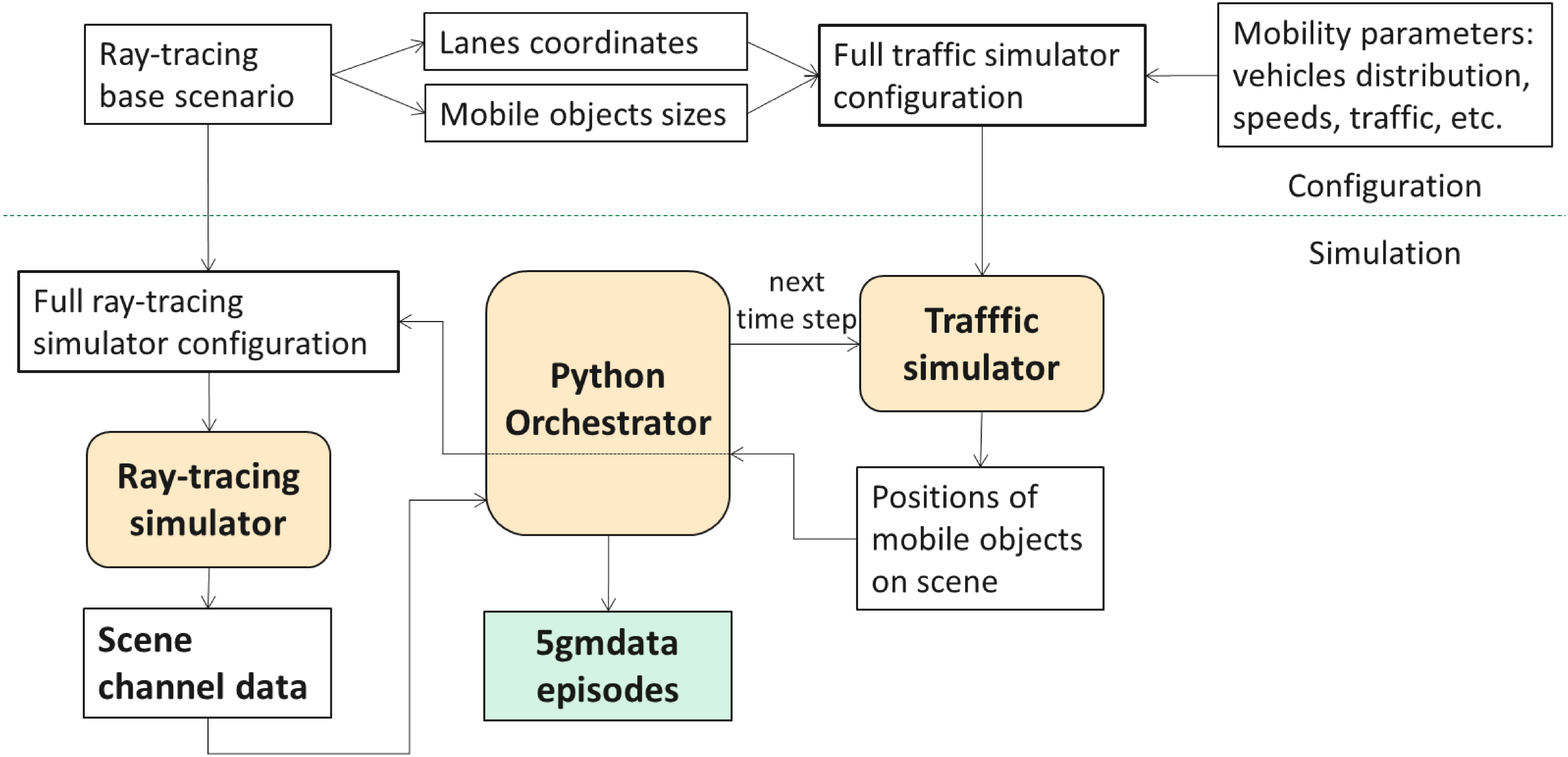}
\caption{Methodology that integrates ray-tracing and traffic simulators.\label{fig:methodology}}
\end{figure}

To support our methodology, we wrote a Python \emph{orchestrator} code to repeatedly invoke the traffic simulator, convert the vehicles position
to a format that can be interpreted by the RT simulator, invoke the latter and post-processing
the RT results to create episodes, as depicted in Fig.~\ref{fig:methodology}. 


\begin{figure}[htb]
\centering
\includegraphics[width=\columnwidth]{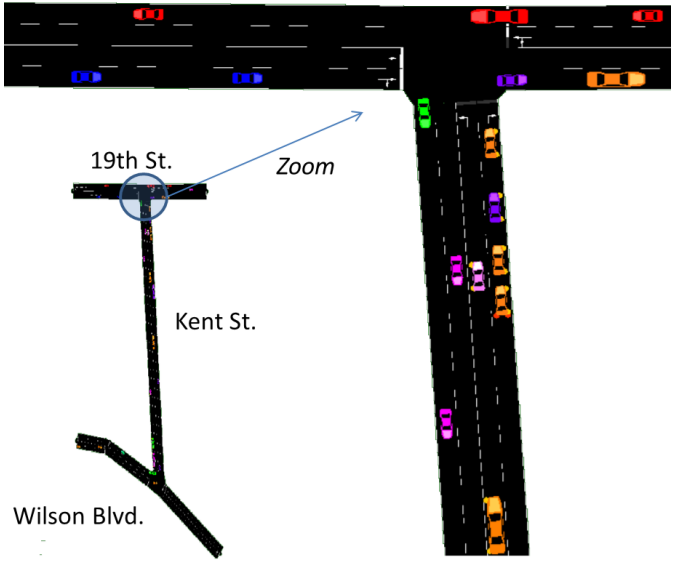}
\caption{Streets of interest for RT in Fig.~\ref{fig:vehicles} as represented in the
traffic simulator. The junction at the right corresponds to a zoom of the intersection
between Kent and 19th streets.\label{fig:sumo}}
\end{figure}

%
The main steps of the proposed methodology can be organized into \emph{configuration} and \emph{simulation} stages. 
In the configuration stage (upper blocks in Fig.~\ref{fig:methodology}), the user provides, e.\,g., information to enable conversion of coordinates between the two main softwares. Fig.~\ref{fig:sumo} illustrates how the streets of interest in 
Fig.~\ref{fig:vehicles} are represented in the traffic simulator after having the
coordinates properly converted. To facilitate the interaction with the
traffic simulator, the orchestrator associates each mobile transmitter or receiver to a mobile object (MOBJ). A MOBJ can also simply play the role of a blocker or scatterer, with no associated transceiver.
In the configuration stage, for each episode, the user specifies the \emph{base scenario}  files. The base scenario files, together with positions for all MOBJs, specified by the traffic simulator, compose all information required for a complete RT simulation. For simplicity, it is assumed in this paper that all episodes are generated with the same base scenario. 

In the simulation stage, the orchestrator invokes the traffic simulator and then positions the MOBJs to compose a scene. Based on the output of the traffic simulator, some files of the base scenario are modified and stored in a unique folder. For each scene, this folder path is stored 
in order to allow reproducing the RT simulation of that scene. This enables the user to later extract additional information through
customized software routines, as well as visualize results for a scene using the RT and traffic simulator's GUIs. Similarly, the corresponding information about the traffic simulation is recorded. For instance, this allows to retrieve the positions $(x,y,z)$ 
and dimensions $(l,w,h)$ of all MOBJs in a given time instant. 

The steps of our methodology are summarized as follows:\\
\textbf{Configuration:}
		\begin{itemize}
			\item \emph{Ray-tracing simulator}
				\begin{itemize}
					\item Define base scenario ``city'', ``terrain'', etc. importing from geographic information systems (GIS) or using computer-aided design (CAD) software
					\item Specify coordinates for RT ``study area'' and ``mobility lanes'' for cars, pedestrians, etc.
					\item Create 3-d MOBJs and specify their electromagnetic properties
				\end{itemize}
			\item \emph{Traffic simulator}
				\begin{itemize}
					\item Import from ray-tracing configuration the lanes coordinates and sizes of MOBJs
					\item Specify MOBJs distribution, routes, traffic statistics, maximum speeds,
					accelerations, etc.
				\end{itemize}
		\end{itemize}
\textbf{Simulation:}
\begin{itemize}
	\item \emph{Python orchestrator code}, repeatedly:
		\begin{itemize}
			\item Randomly selects the start of an episode
			\item With sampling period $T_{\textrm{sam}}$, for each scene:
			\begin{itemize}
			\item Invoke the traffic simulator and get the positions of all MOBJs
			\item Create a full configuration for the ray-tracing simulator and execute it
			\end{itemize}
			\item Retrieve the ray-tracing outputs and organize them as episodes.
		\end{itemize}
\end{itemize}
~\\


Following the steps of our methodology lead to the creation of simulation data of 5G \textbf{mm}Wave \textbf{M}I\textbf{M}O systems involving \textbf{m}obility (or \emph{5GMdata}) that can be used in different applications. In summary, the dataset stores for each episode: base scenario folder and traffic simulator configuration file paths, episode start time, sampling period $T_{\textrm{sam}}$, number of transmitters, receivers and MOBJs, dimensions of all MOBJs, mappings between transmitters / receivers and MOBJs, coordinates of the \emph{RT study area} and number $L$ of rays per transmitter / receiver pair. Besides, the episode contains information for all its scenes.
For a given scene, the information collected from the outputs of the RT and traffic simulators for each transmitter / receiver pair $(m,n)$ are: average time of arrival $\overline \tau_{mn}$ (the subscripts $mn$ will be omitted hereafter), total transmitted $\hat P_\textrm{tx}$ and received $\hat P_\textrm{rx}$ powers, and
for the $\ell$-th ray, $\ell=1,\ldots,L$, complex channel gain $\alpha_{\ell}$, time of arrival $\tau_\ell$, angles $\phi_\ell^D$, $\theta_\ell^D$, $\phi_\ell^A$, $\theta_\ell^A$, corresponding respectively to azimuth and elevation for departure and arrival. Besides, a string $s_\ell$ stores all
ray ``interactions'' (reflection, diffraction, DS) and facilitates distinguishing LOS and NLOS situations.

After 5GMdata is obtained, additional post-processing stages can generate the required data for  specific target applications.
For concreteness, the next section discusses possible uses of 5GMdata in the V2I  context.

\section{Machine Learning for Beam-Selection in V2I}
\label{sec:beam_selection}



In this section, we illustrate the application of our proposed framework. Specifically, we generate data for the application of ML to predict the best beam pairs in the context of mmWave to cellular systems (the V2I setting).

\subsection{Brief literature review of beam-selection}

MmWave MIMO 
is a means to exchange sensor data in vehicular systems~\cite{Prelcic17}.
A main challenge is that mmWave, as initially envisioned for this application, requires the pointing of narrow beams at both the transmitter and receiver. 
Taking into account extra information such as out-of-band measurements and vehicles positions can 
reduce the time needed to find the best beam pair~\cite{ali_millimeter_2017,va_inverse_2017,Prelcic17}. 
Beam training is part of standards such as IEEE 802.11ad and 5G, and has also been extensively studied in the context of wireless personal and local area networks (see, e.\,g.~\cite{kim_fast_2014,zhou_enhanced_2017} and references therein).
Among the several problems related to beam training and tracking in distinct scenarios~\cite{va_inverse_2017}, this paper focuses on a subset collectively called \emph{beam-selection} in V2I.
The goal is to choose the best pair of beams for analog beamforming, with both transmitter and receiver having antenna arrays with only one radio frequency (RF) chain and fixed beam codebooks. The next subsection describes the information extracted for V2I ML experiments.

\subsection{Dataset for machine learning in V2I}

Using our methodology, 5GMdata is organized with the following characteristics. We 
generated all episodes with the base scenario depicted in
Fig.~\ref{fig:vehicles}, which corresponds to a 3-d model that is part of Wireless InSite's examples. The scenario represents a region of Rosslyn,\footnote{The lanes adopted in this paper follow the 3-d model geometry but do not actually exist.} Virginia, which was studied e.\,g. in~\cite{kim_radio_1999}.
The RT area of study is a rectangle of approximately $337 \times 202 \textrm{~m}^2$.
 A transmitter is located at the RSU on Kent Street, as depicted in Fig.~\ref{fig:vehicles}.
We placed receivers on top of 10 vehicles (some identified in red in Fig.~\ref{fig:vehicles}) and obtained 50 scenes per episode. The experiments reported in this paper concern 116 episodes.
Wireless InSite's command-line
\emph{wibatch} is adopted for its support to the X3D model, which implements DS. 
Table~\ref{tab:simulationParameters} describes the most important simulation parameters.

\begin{table}[htb]
\centering
\caption{Simulation parameters.\label{tab:simulationParameters}}
\scalebox{1}{
\begin{tabular}{| c | c | }
\hline
\multicolumn{2}{|c|}{ \textbf{Ray-tracing parameters }} \\ \hline
Carrier frequency &  60 GHz  \\ \hline
RSU transmitted power & 0 dBm \\ \hline
RSU antenna height & 5 m \\ \hline
Antenna (Tx and Rx) & Half-wave dipole \\ \hline
Propagation model & X3D \\ \hline
Terrain and city material & ITU concrete 60 GHz \\ \hline
Vehicle material & Metal \\ \hline
Ray spacing (degrees) & 1 \\ \hline
Num. $L$ of strongest rays & 25  \\ \hline
Diffuse scattering model & Lambertian  \\ \hline
DS max. reflections ($N_{\textrm{max}}^{\textrm{DS}}$) & 2  \\ \hline
DS coefficients ($S$) & 0.4 (concrete), 0.2 (metal) \\ \hline
\multicolumn{2}{|c|}{ \textbf{Traffic parameters }} \\ \hline
Number of lanes & 4 \\ \hline
Vehicles & car, truck, bus \\ \hline
Lengths, respectively (m) & 4.645, 12.5, 9.0  \\ \hline
Heights, respectively (m) & 1.59, 4.3, 3.2  \\ \hline
Probabilities, respectively & 0.7, 0.1, 0.2 \\ \hline
Average speed (m/s) & 8.2 \\ \hline
Sampling period $T_{\textrm{sam}}$ (s) & 0.1 \\ \hline
\end{tabular}}
\end{table}

After the 5GMdata is obtained, the following post-processing is adopted. 

\subsection{Machine learning input features}
\label{sec:input_features}
The ML problems illustrated in this paper address beam-selection based
solely on vehicles positions and sizes. It is assumed that the RSU receives
through an error-free channel the position and a unique index of all vehicles 
for each scene. 
Based on the vehicle index, the RSU knows its
dimension and may incorporate it as extra out-of-band information provided
to the ML algorithm. The position and identity information is then represented
as a matrix. Based on this input, a ML algorithm should estimate parameters
of interest to beam-selection.

\begin{figure}[htb]
\centering
\includegraphics[width=2cm]{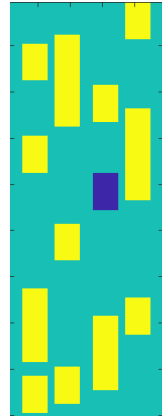}
\caption{Image corresponding to an example of an input features matrix $\mathbf{Q}_s$ representing
13 vehicles in four (vertical) lanes. The elements of $\mathbf{Q}_s$ corresponding to the pixels of the receiver (blue vehicle) are $+1$ while elements with value $-1$ (yellow) identify the positions of the other vehicles.\label{fig:features}}
\end{figure}

The \emph{V2I study area}, with $23 \times 250 \textrm{~m}^2$, is a subarea of the RT study area consisting basically of the street in which the RSU is located.
A grid with resolution of $1 \times 1 \textrm{~m}^2$ is adopted to represent the V2I study area, leading to a matrix $\mathbf{Q}_s$ of dimension $23 \times 250$ to represent each scene $s$.
The RSU has a fixed position, which is then not explicitly specified in $\mathbf{Q}_s$.
A negative element in $\mathbf{Q}_s$ indicates that the corresponding location is occupied (even partially) by a vehicle that is not a receiver or transmitter. The magnitude of this negative value indicates
    the vehicle's height. For example, $-1$ and $-2$ represent a car and a truck, respectively.
A positive integer value $r$ in $\mathbf{Q}_s$ represents the location of the $r$-th receiver, while 0 denotes the position is not occupied. 
Fig.~\ref{fig:features} illustrates an example where the receiver is blue and the surrounding vehicles are yellow.

The next subsection describes post-processing schemes to extract information useful for generating
the outputs of ML problems.

\subsection{Post-processing ray-tracing outputs}
\label{sec:postprocessing}

The definition of desired beam-selection outcomes requires to model the composition of channels
and beams based on the RT outputs. 
The estimated beams allow, for instance, the definition of \emph{optimal} beam pair indices to be used as target outputs (or \emph{labels}) in \emph{supervised learning}~\cite{geron2017hands}.
The beams are assumed to have a beamwidth of $\beta$ radians.
We consider two different channel models in the next paragraphs, but others can be adopted such as 
\emph{wideband} models (see, e.\,g.,~\cite{heath_overview_2016} and references therein).
The first case is called \emph{mmWave massive MIMO} model and considers each ray as a beam.
%
%
The other case represents a more realistic situation that assumes fixed beam codebooks at  transmitter and receiver, and uses basic signal processing techniques to obtain the received power for beam pairs.

\subsubsection{MmWave Massive MIMO}

In the massive MIMO case, the number $N$ of antenna elements is large, and assuming $N\to\infty$ implies a small beamwidth $\beta \to 0$. Therefore, in this model, the departure $[\phi_{\ell_*}^D, \theta_{\ell_*}^D]$ and $[\phi_{\ell_*}^A, \theta_{\ell_*}^A]$ arrival directions of the \emph{strongest} ray $\ell_*$ indicate the target \emph{optimal} angles, which can be used in regression
problems.
In the case of interest in classification problems, one can quantize the angles $[\phi_{\ell_*}^D, \theta_{\ell_*}^D,\phi_{\ell_*}^A, \theta_{\ell_*}^A]$ using vector or scalar quantization. If the latter is used, the angles can be quantized into four indices $[D_\text{azi}, D_\text{ele}, A_\text{azi}, A_\text{ele}]$ according to their dynamic ranges in the training set.
These indices can be 
eventually converted to a single label for traditional classification. Typically, due to the scenario geometry, the number $M$ 
of unique vectors that occur in the dataset is smaller than the total number of  Cartesian products among $[D_\text{azi}, D_\text{ele}, A_\text{azi}, A_\text{ele}]$. It is therefore useful to pre-process 
the quantized values and map the vectors that actually appeared in the data into 
the range $\{1, 2, \cdots,M\}$, where $M$ is the number of \emph{class labels}. When training
classifiers, one can then conveniently represent the labels with \emph{one-hot} encoding to facilitate
training neural networks, for example~\cite{geron2017hands}. 


\subsubsection{Codebook-based beams}
In practical mmWave systems, $N$ is finite and influences the beamwidth $\beta_\mathrm N > 0$ for the projected beam given the antenna arrays. 
To take this in account, one can estimate the MIMO channel by combining the RT output with the mmWave \emph{geometric channel model} as follows (see, e.\,g.,~\cite{va_inverse_2017} and references therein):
\begin{align}
\mathbf{H}_{mn} = \sqrt{N_t N_r}\sum_{\ell = 1}^L \alpha_{\ell} \mathbf{a}_r(\phi_\ell^A, \theta_\ell^A)\mathbf{a}^*_t(\phi_\ell^D, \theta_\ell^D), 
\end{align}
where $N_t$ and $N_r$ are the numbers of antennas at the $n$-th transmitter and $m$-th receiver, $\alpha_\ell$ is the complex channel gain, $\mathbf{a}_r(\phi_\ell^A, \theta_\ell^A)$ and $\mathbf{a}^*_t(\phi_\ell^D, \theta_\ell^D) $ are the steering vectors at the receiver and transmitter for the $\ell$-th path, respectively. 
We also assume DFT codebooks $\mathcal{C}_t = \{\bar{\mathbf{w}}_1, \cdots, \bar{\mathbf{w}}_{|\mathcal{C}_t|}\}$ and $\mathcal{C}_r= \{\bar{\mathbf{f}_1}, \cdots, \bar{\mathbf{f}}_{|\mathcal{C}_r|}\}$ at the transmitter and the receiver sides, where $|\mathcal{C}_t|$ and $|\mathcal{C}_r|$ are the cardinalities of these codebooks correspondingly. In specific, we have $|\mathcal{C}_t| = N_t = |\mathcal{C}_r| = N_r$ in our case. The beam pair $[p,q]$ is converted into a unique index $i \in \{1, 2, \cdots,M\}$, where $M \le  |\mathcal{C}_t||\mathcal{C}_r|$. For the $i$-th pair, the received signal can be calculated as
\begin{align}
\label{eq:beamOutput}
y_i= \mathbf{w}_i^* \mathbf{H} \mathbf{f}_i
\end{align}
and the \emph{optimal} beam pair index $\hat i$ is given by
\begin{align}
\hat i = \arg\max_{i\in\{1, \cdots, M\}} |y_i|.
\end{align}

This post-processing and data representations allow the formulation of several ML
problems for beam-selection. Some alternatives are discussed in the next section.

\section{Examples of Experiments with 5GMdata}
\label{sec:experiments}

Next we illustrate some ML experiments with the described dataset.
 Three examples of machine learning problems are described.
Only the third example uses the time evolution while the others are \emph{drop-based} and depend only on data from a given scene.
In any ML problem, care must be exercised to use an evaluation strategy that allows to estimate
the generalization capability of the classifiers. 
For example, when splitting the dataset into disjoint training, validation and test sets, we shuffle the episodes and not the scenes, given that scenes are similar along an episode.
There are many other details in elaborate simulations and, to promote reproducibility, the dataset and associated code will be made available 
at~\cite{5GMdata}. 

\subsection{Conventional drop-based classification}
\label{sec:drop-classification}
We pose the beam-selection as a classification task in which the target output is the best beam pair index $\hat i$. The input features correspond to the matrix described in Section~\ref{sec:input_features} with the following modification: we generate $\mathbf{Q}_{s,r}$, a modified version of $\mathbf{Q}_s$ for each receiver $r$, assuming a value $+1$ for all $\mathbf{Q}_s$ elements corresponding to the target receiver $r$, while all other receivers in the given scene $s$ are represented with $-1$ (instead of their original positive values in $\mathbf{Q}_s$). The 116 episodes (with 50 scenes each) are split and 34 episodes used for testing. For each receiver that is part of a given scene, a classification \emph{example} is obtained, leading to a total of 41,023 examples for training and test. In 16,977 cases, the receiver is in the (larger) RT study area but not in the V2I study area.
Among the examples, there is LOS in 25,174 cases and NLOS in 15,849. Transmitter and receivers 
had $4 \times 4$ uniform planar antenna arrays (UPA), such that $N_t = N_r = 16$. There are $M=61$ classes (optimum beam pairs) among the possible $|\mathcal{C}_t||\mathcal{C}_r| = 256$ pairs.
Table~\ref{tab:classification} presents the accuracy using this data for distinct learning algorithms.
The hyperparameters and other details can be obtained at~\cite{5GMdata}.

\begin{table}[htb]
\centering
\caption{Classification results.\label{tab:classification}}
\scalebox{1}{
\begin{tabular}{ c | c | c |}
\cline{2-3} 
& \multicolumn{2}{c|}{Accuracy (\%)}\\  \hline
\multicolumn{1}{|c|} {Classifier} & All data & Only NLOS\\ \hline
\multicolumn{1}{|c|} {Linear SVM} & 33.2 &  12.4 \\ \hline
\multicolumn{1}{|c|} {AdaBoost} & 55.0 &  22.5 \\ \hline
\multicolumn{1}{|c|} {Decision tree} & 55.5 & 27.3 \\ \hline
\multicolumn{1}{|c|} {Random Forest}  & 63.2 &  36.9 \\ \hline
\multicolumn{1}{|c|} {Deep neural network} & 63.8 &  38.1 \\ \hline
\end{tabular}}
\end{table}

The data used for the ``All data'' column in Table~\ref{tab:classification} had approximately 60\% of examples in LOS and this is comparable to the maximum accuracy of 63\%. The LOS case could be
addressed with simple geometry.
Restricting attention only to NLOS examples leads to the results in the right column of Table~\ref{tab:classification}. In spite of not being our goal to investigate performance levels
with this relatively small amount of data, the results indicate that deep learning 
has clear advantage over the other tested methods. While deep neural networks are more popular,
random forest are also ``deep'' in the sense that consist of an ensemble (obtained with \emph{bagging})
of decision trees~\cite{geron2017hands}.


Some classifiers in Table~\ref{tab:classification} reach zero errors on the training set while
the errors in the test set are relatively large. Such \emph{overfitting} indicates that more data is needed to avoid
evaluating deep learning algorithms solely on \emph{small data} regime.

\subsection{Conventional drop-based multivariate regression}

The 5GMdata can also be used
for regression tasks. For example, estimating the
angles of departure and arrival in beam-selection can be
cast as a multivariate regression problem in which the desired output is 
$[\phi_{\ell_*}^D, \theta_{\ell_*}^D, \phi_{\ell_*}^A, \theta_{\ell_*}^A]$ and the 
input is the matrix previously used.
Table~\ref{tab:regression} presents the root mean-squared error (RMSE) for estimations
with neural networks in this problem. The deep architecture outperforms the shallow
(with only one hidden layer), but in both cases, the deviations from the target
 are relatively large, especially for the azimuth angles.

\begin{table}[htb]
\centering
\caption{Regression results in terms of RMSE (angles in degrees).\label{tab:regression}}
\scalebox{1}{
\begin{tabular}{c|c|c|c|c|}
\cline{2-5} 
& \multicolumn{2}{c|}{ Departure (Tx)} & \multicolumn{2}{c|}{ Arrival (Rx)}\\  \hline
\multicolumn{1}{|c|} {Regressor} &   Ele. & Azi. &  Ele. & Azi. \\ \hline
\multicolumn{1}{ |c| } {Shallow neural network} & 6.5 &  137.4 &     7.9 &  180.6 \\ \hline
\multicolumn{1}{ |c| } {Deep neural network} &  4.8 &   49.9 & 6.2 & 102.8 \\ \hline
\end{tabular}}
\end{table}

In some applications, it is not essential to find the optimum beam direction, but grant an overall quality of service. The next subsection presents an example that uses regression
to estimate the output powers of each beam within a reinforcement learning setup.

\subsection{Deep reinforcement learning}

One of the requirements of our methodology is to provide the history of the channel evolution over time. The time evolution facilitates, for instance, taking into account 
the interaction of the mmWave PHY with the media access control (MAC) and upper layers.
Designing the mmWave MAC 
is an important issue for 5G~\cite{dutta_frame_2017} and ML can be useful in this context. The next paragraphs aim at giving a concrete example within the framework of \emph{deep reinforcement learning}~\cite{geron2017hands}. Again, the goal is not to outperform previously published methods, but
illustrate how 5GMdata can be effectively used.

\begin{figure}[htb]
\centering
\includegraphics[width=6.5cm]{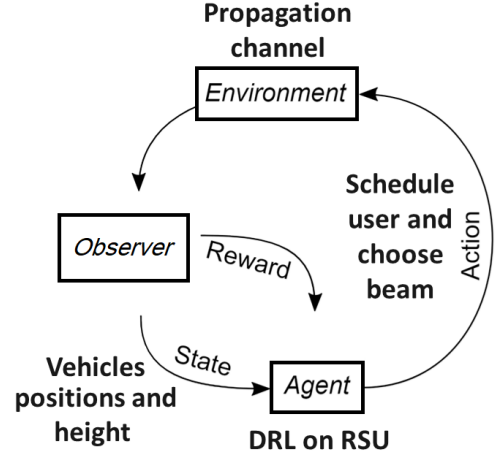}
\caption{Deep reinforcement learning for multi-user beam-selection in V2I.\label{fig:drl}}
\end{figure}

In reinforcement learning~\cite{geron2017hands}, an \emph{agent} is capable of \emph{actions} that influence the \emph{state} of the \emph{environment} to obtain a \emph{reward}. 
Deep reinforcement learning (DRL) is often associated to having
a deep neural network to choose the actions.
Fig.~\ref{fig:drl} indicates how beam-selection can be cast as a DRL problem.
In this case, the analog beamforming architecture is assumed and the action is to schedule a user in each time-slot of duration
$T_{\textrm{sam}}$, together with its beam pair index $i$. The state at scene $s$ is represented by the  matrix $\mathbf{Q}_s$ described in Section~\ref{sec:input_features}, which represents
all receivers of interest.
There is flexibility on choosing the reward. For example, the reward can be based on figures of merit such
as throughput, fairness, energy or their combination. Different from conventional  regression or classification problems, 
the agent does not seek to find the optimum solution at each time instant, but properly allocate resources among users to
maximize the reward over time.

For simplicity, the rewards are based on the value $z_{s,r,i} = 20 \log_{10}|y_{s,r,i}|$ for scene $s$, receiver $r$ and beam pair $i$, where $y_{s,r,i}$ is given by Eq.~(\ref{eq:beamOutput}). To speed up convergence, the extreme values 
$z_{\min}=\min_{r,i} z_{s,r,i}$ and $z_{\max}=\max_{r,i}z_{s,r,i}$ for each scene $s$ are used to obtain
 $\overline z_{s,r,i} = (z_{s,r,i} - z_{\min})/(z_{\max}-z_{\min})$ in the range $[0,1]$.
Improved numerical stability is achieved by using a floor value for $z_{\min}$, such as $z_{\max}-200$.
Assuming there are $N_{\textrm{sce}}$ scenes per episode, the agent distributes $N_{\textrm{sce}}$ time-slots among the receivers and chooses the beam pair for each one.
These decisions are represented by arrays $\textbf{r}$ and $\textbf{i}$, whose elements inform the chosen receiver $\textbf{r}[s]$ and beam pair $\textbf{i}[s]$ at scene $s$, respectively.
The reward is then $r_s = \overline z_{s,\textbf{r}[s],\textbf{i}[s]}$ unless there is an \emph{outage}.
The user $r$ is in outage when the agent does not allocate time-slots for $r$ over $N_{\textrm{out}}$ or more consecutive scenes. The reward is $r_s = r_{\textrm{out}}$ if there is any user in outage at scene $s$. The value of $r_{\textrm{out}}$ is typically a negative number to penalize the outage occurrence.
The average reward per scene for an episode $e$ is then $R_e = (1/N_{\textrm{sce}}) \sum_{s=1}^{N_{\textrm{sce}}} r_s$.
If there is no chance of an outage ($N_{\textrm{out}} \to \infty$)
and assuming an agent capable of always choosing the receiver with the strongest power $z_{s,r,i}$, the average reward would be $R_e = 1, \forall e$, given
the way $\overline z_{s,r,i}$ is defined.

To model this problem as DRL, we adopt a cascade of two networks. The first is a convolutional deep neural network ${\cal{N}}_1$ that has $\mathbf{Q}_{s}$ as its input and outputs estimates of $\overline z_{s,r,i}$ for scene $s$, organized as an array with $N_{\textrm{rec}} \times |\mathcal{C}_t||\mathcal{C}_r|$ elements.
The peak value for each row of this array indicate the best beam pair per receiver. This array is part of the input to the second network ${\cal{N}}_2$, which is also composed by
an array of binary elements that indicate the time-slots allocated to receivers over the previous $N_{\textrm{out}}$ scenes. The network ${\cal{N}}_2$ has $N_{\textrm{rec}}$ outputs, 
for each scene $s$, the receiver that should be allocated to the corresponding time-slot.
The output of ${\cal{N}}_1$ is then used to choose the beam pair for the chosen receiver.


Regarding the input to the first network ${\cal{N}}_1$, 
the matrix $\mathbf{Q}_s$ representing all receivers is converted into $N_{\textrm{rec}}$ matrices $\mathbf{Q}_{s,r}$, that inform where a given receiver $r$ is located, while treating all other receivers as regular vehicles (turning their corresponding positive values in $\mathbf{Q}_{s}$ into $-1$), similar to the scheme used in Section~\ref{sec:drop-classification}. The  input to ${\cal{N}}_1$ is then the concatenation of $N_{\textrm{rec}}$ matrices $\mathbf{Q}_{s,r}$. Conceptually, it would 
be possible to build for each receiver $r_*$, a sub-network with input $\mathbf{Q}_{s,r_*}$ that outputs estimates of $\overline z_{s,r_*,i}$. To decrease the
computational cost, instead, the layers are shared among the distinct $\mathbf{Q}_{s,r}$ (sharing layers
is a feature of DL packages such as Keras~\cite{chollet2015keras}). Another speed up is
obtained by training ${\cal{N}}_1$ using supervised regression to estimate 
the outputs $\overline z_{s,r_*,i}, \forall i$, given the corresponding inputs $\mathbf{Q}_{s,r_*}$,  $r_*=1,\ldots,N_{\textrm{rec}}$, 
A third aspect is that after an action in RL, the environment state needs to be updated accordingly. In our case, it would be inconvenient to execute the RT and 
traffic simulators within the DRL loop. 
In this experiment however, invoking the simulators is not necessary given that their pre-calculated outputs suffice to obtain $\overline z_{s,r,i}$.

\begin{figure}[htb]
\centering
\includegraphics[width=\columnwidth]{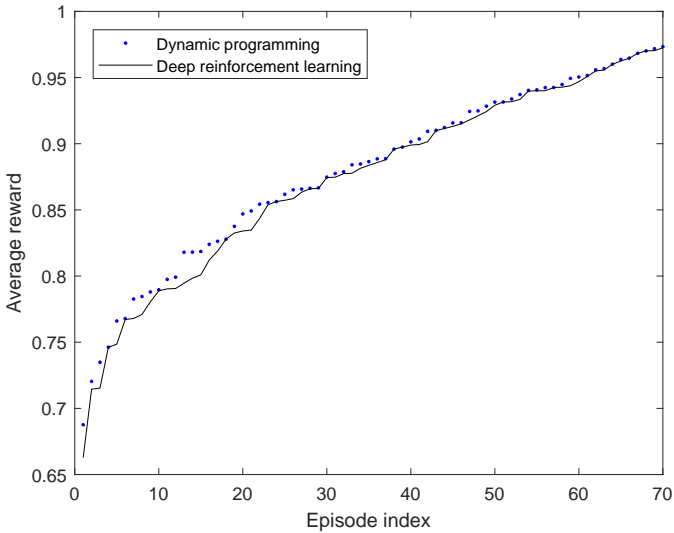}
\caption{Performance of DRL in the training set and comparison with the optimum allocation
obtained via dynamic programming.\label{fig:drl_comparison}}
\end{figure}

The Deep Q Learning (DQN) algorithm is used, as implemented in Keras-RL~\cite{plappert2016kerasrl}.
It is assumed $r_{\textrm{out}}=-3$, $N_{\textrm{out}}=3$, $N_{\textrm{rec}}=2$ 
and $4 \times 4$ UPAs for both transmitter and receivers. Seventy episodes are used 
in experiments to illustrate the learning process. 
The generalization capability is not evaluated in this paper.
Considering first the performance of ${\cal{N}}_1$ in the (embedded) supervised regression task, it achieved an average $\textrm{RMSE} = 0.074$ while estimating $\overline z_{s,r,i}$.
Still in a supervised learning settings, an accuracy of 67.5\% is obtained when the strongest beam pair indicated by the estimates of $\overline z_{s,r,i}$ is used as a classifier outputs.
The performance of the overall DRL model (cascade of ${\cal{N}}_1$ and ${\cal{N}}_2$) is presented in
Fig.~\ref{fig:drl_comparison}, which shows the average rewards in all episodes. For comparison,
assuming the regression values estimated by ${\cal{N}}_1$, Fig.~\ref{fig:drl_comparison} also
indicates the optimum time-slot allocation 
obtained with dynamic programming. The average reward over all episodes for dynamic programming in this case and DRL are 0.879 and 0.874, respectively. If the actual values of $\overline z_{s,r,i}$
are passed to the dynamic programming routine (instead of ${\cal{N}}_1$ estimates), the average reward increases
to 0.891. These results indicate that the DRL is able to learn the task of simultaneously allocating receivers to time-slots and choosing beam pairs.




\section{Conclusion}
\label{sec:conclusion}

This paper presented a methodology for generating 5G propagation channel data
that decouples the tasks of modeling mobility and channel. This facilitates
the use of advanced features of traffic simulators. Given the current lack of freely available large amount of data for benchmarking
deep learning algorithms in 5G, it is reasonable to use simulations especially
in complicated configurations. 
The generated data incorporates the channel
evolution over time and can be used, for example, in machine learning problems involving aspects of the
5G PHY with constraints from MAC and upper
layers.
The focus was
mmWave MIMO but the methodology can be used in other scenarios.
Future work includes simulating different sites and scenarios, while validating
some of them with measurements.
Currently, it is not clear how detailed must be the description of 
ray-tracing scenarios to support broad conclusions such as average performance
on distinct sites. Measurements can help tuning the methodology. Besides
accurate modeling, it is important to minimize the computational cost.
An alternative to speed up simulations is to combine ray-tracing outputs
with statistical models and eventually avoid the longer simulation time
caused by the diffuse-scattering feature. After escaping the small
data regime, deep learning in 5G can be investigated using a systematic
and reproducible experimental procedure.

%
%
%
\section*{Acknowledgment}
The work of A. Klautau and P. Batista was supported in part by CNPq, Brazil (processs 201493/2017-9/PDE), and of
R. Heath and Y. Wang by the U.S. Department of Transportation through the Data-Supported Transportation Operations and Planning Tier 1 University Transportation Center, 
in part by the National Science Foundation under Grant 1711702, and by a gift from Huawei.
The work of N. González-Prelcic was supported by the Spanish Government and the European Regional Development Fund through the Project MYRADA under Grant TEC2016-75103-C2-2-R.


\bibliographystyle{IEEEtran}
\bibliography{IEEEabrv,refer,ak_references,zotero_exported_items_processed}

\begin{thebibliography}{10}
\providecommand{\url}[1]{#1}
\csname url@samestyle\endcsname
\providecommand{\newblock}{\relax}
\providecommand{\bibinfo}[2]{#2}
\providecommand{\BIBentrySTDinterwordspacing}{\spaceskip=0pt\relax}
\providecommand{\BIBentryALTinterwordstretchfactor}{4}
\providecommand{\BIBentryALTinterwordspacing}{\spaceskip=\fontdimen2\font plus
\BIBentryALTinterwordstretchfactor\fontdimen3\font minus
  \fontdimen4\font\relax}
\providecommand{\BIBforeignlanguage}[2]{{%
\expandafter\ifx\csname l@#1\endcsname\relax
\typeout{** WARNING: IEEEtran.bst: No hyphenation pattern has been}%
\typeout{** loaded for the language `#1'. Using the pattern for}%
\typeout{** the default language instead.}%
\else
\language=\csname l@#1\endcsname
\fi
#2}}
\providecommand{\BIBdecl}{\relax}
\BIBdecl

\bibitem{jiang_machine_2017}
C.~Jiang, H.~Zhang, Y.~Ren, Z.~Han, K.~C. Chen, and L.~Hanzo, ``Machine
  {Learning} {Paradigms} for {Next}-{Generation} {Wireless} {Networks},''
  \emph{{IEEE} Wireless Commun.}, vol.~24, no.~2, pp. 98--105, Apr. 2017.

\bibitem{klaine_survey_2017}
P.~V. Klaine, M.~A. Imran, O.~Onireti, and R.~D. Souza, ``A {Survey} of
  {Machine} {Learning} {Techniques} {Applied} to {Self}-{Organizing} {Cellular}
  {Networks},'' \emph{{IEEE} Commun. Surveys Tuts.}, vol.~19, no.~4, pp.
  2392--2431, 2017.

\bibitem{Lecun2015}
Y.~LeCun, Y.~Bengio, and G.~Hinton, ``Deep learning,'' \emph{Nature}, vol. 521,
  pp. 436--444, 2015.

\bibitem{oshea_introduction_2017}
T.~O’Shea and J.~Hoydis, ``An {Introduction} to {Deep} {Learning} for the
  {Physical} {Layer},'' \emph{{IEEE} Trans. on Cogn. Commun. Netw.}, vol.~3,
  no.~4, pp. 563--575, 2017.

\bibitem{he_integrated_2017}
T.~Y. He, N.~Zhao, and H.~Yin, ``Integrated {Networking}, {Caching} and
  {Computing} for {Connected} {Vehicles}: {A} {Deep} {Reinforcement} {Learning}
  {Approach},'' \emph{{IEEE} Trans. Veh. Technol.}, vol.~67, no.~1, pp. 44--55,
  2018.

\bibitem{han_two-dimensional_2017}
G.~Han, L.~Xiao, and H.~V. Poor, ``Two-dimensional anti-jamming communication
  based on deep reinforcement learning,'' in \emph{Proc. IEEE Int. Conf.
  Acoust. Speech Signal Proc. ({ICASSP})}, Mar. 2017, pp. 2087--2091.

\bibitem{west_deep_2017}
N.~E. West and T.~O'Shea, ``Deep architectures for modulation recognition,'' in
  \emph{Proc. IEEE Int. Symp. Dynamic Spectrum Access Networks (DySPAN)}, Mar.
  2017, pp. 1--6.

\bibitem{atallah_deep_2017}
R.~Atallah, C.~Assi, and M.~Khabbaz, ``Deep reinforcement learning-based
  scheduling for roadside communication networks,'' in \emph{Proc. Int. Symp.
  {Modeling} {Optimization} {Mobile}, {Ad} {Hoc}, {Wireless} {Networks}
  ({WiOpt})}, May 2017, pp. 1--8.

\bibitem{atallah_reinforcement_2017}
R.~F. Atallah, C.~M. Assi, and J.~Y. Yu, ``A {Reinforcement} {Learning}
  {Technique} for {Optimizing} {Downlink} {Scheduling} in an {Energy}-{Limited}
  {Vehicular} {Network},'' \emph{{IEEE} Trans. Veh. Technol.}, vol.~66, no.~6,
  pp. 4592--4601, Jun. 2017.

\bibitem{xu_deep_2017}
Z.~Xu, Y.~Wang, J.~Tang, J.~Wang, and M.~C. Gursoy, ``A deep reinforcement
  learning based framework for power-efficient resource allocation in cloud
  {RANs},'' in \emph{Proc. IEEE Int. Conf. {Communications} ({ICC})}, May 2017,
  pp. 1--6.

\bibitem{ferreira_multi-objective_2017}
P.~V.~R. Ferreira, R.~Paffenroth, A.~M. Wyglinski, T.~M. Hackett, S.~G. Bilén,
  R.~C. Reinhart, and D.~J. Mortensen, ``Multi-objective reinforcement
  learning-based deep neural networks for cognitive space communications,'' in
  \emph{Proc. {Cognitive} {Communications} {Aerospace} {Applications}
  {Workshop} ({CCAA})}, Jun. 2017, pp. 1--8.

\bibitem{Ekman2017}
``Machine learning for beam based mobility optimization in {NR},''
  \url{http://www.diva-portal.org/smash/get/diva2:1088857/FULLTEXT01.pdf},
  {Björn} {Ekman}, Master of Science Thesis, Linköping University, 2017.

\bibitem{Shen2018}
J.~S. et~al, ``Natural {TTS} synthesis by conditioning {WaveNet} on mel
  spectrogram predictions,'' in \emph{https://arxiv.org/abs/1712.05884}, 2018.

\bibitem{maccartney_flexible_2017}
G.~R. MacCartney, H.~Yan, S.~Sun, and T.~S. Rappaport, ``A flexible wideband
  millimeter-wave channel sounder with local area and {NLOS} to {LOS}
  transition measurements,'' in \emph{Proc. IEEE Int. Conf. {Communications}
  ({ICC})}, May 2017, pp. 1--7.

\bibitem{heath_overview_2016}
R.~W. Heath, N.~González-Prelcic, S.~Rangan, W.~Roh, and A.~M. Sayeed, ``An
  {Overview} of {Signal} {Processing} {Techniques} for {Millimeter} {Wave}
  {MIMO} {Systems},'' \emph{{IEEE} J. Sel. Topics Signal Process.}, vol.~10,
  no.~3, pp. 436--453, Apr. 2016.

\bibitem{Rappaport14}
T.~S. Rappaport, R.~W. Heath, R.~C. Daniels, and J.~N. Murdock,
  \emph{Millimeter Wave Wireless Communications}.\hskip 1em plus 0.5em minus
  0.4em\relax Prentice Hall, 2014.

\bibitem{fuschini_ray_2015}
\BIBentryALTinterwordspacing
F.~Fuschini, E.~M. Vitucci, M.~Barbiroli, G.~Falciasecca, and V.~Degli-Esposti,
  ``\BIBforeignlanguage{en}{Ray tracing propagation modeling for future
  small-cell and indoor applications: {A} review of current techniques},''
  \emph{\BIBforeignlanguage{en}{Radio Sci.}}, vol.~50, no.~6, p. 2015RS005659,
  Jun. 2015. [Online]. Available:
  \url{http://onlinelibrary.wiley.com/doi/10.1002/2015RS005659/abstract}
\BIBentrySTDinterwordspacing

\bibitem{rumney_critical_2017}
M.~Rumney, ``The critical importance of accurate channel modelling for the
  success of {mmWave} {5G},'' in \emph{Proc. 11th {European} Conf. {Antennas}
  {Propagation} ({EuCAP})}, Mar. 2017, pp. 3688--3691.

\bibitem{turkka_spatial_2016}
J.~Turkka, P.~Kela, and M.~Costa, ``On the spatial consistency of stochastic
  and map-based {5G} channel models,'' in \emph{{IEEE} Conf. {Standards}
  {Communications} {Networking} ({CSCN})}, Oct. 2016, pp. 1--7.

\bibitem{geron2017hands}
A.~G{\'e}ron, \emph{Hands-On Machine Learning with Scikit-Learn and TensorFlow:
  Concepts, Tools, and Techniques to Build Intelligent Systems}.\hskip 1em plus
  0.5em minus 0.4em\relax O'Reilly Media, 2017.

\bibitem{Wang17}
\BIBentryALTinterwordspacing
T.~Wang, C.~Wen, H.~Wang, F.~Gao, T.~Jiang, and S.~Jin, ``Deep learning for
  wireless physical layer: Opportunities and challenges,'' \emph{CoRR}, vol.
  abs/1710.05312, 2017. [Online]. Available:
  \url{http://arxiv.org/abs/1710.05312}
\BIBentrySTDinterwordspacing

\bibitem{OShea17}
\BIBentryALTinterwordspacing
T.~J. O'Shea, T.~Erpek, and T.~C. Clancy, ``Deep learning based {MIMO}
  communications,'' \emph{CoRR}, vol. abs/1707.07980, 2017. [Online].
  Available: \url{http://arxiv.org/abs/1707.07980}
\BIBentrySTDinterwordspacing

\bibitem{Mismar2018}
F.~B. Mismar and B.~L. Evans, ``Partially blind handovers for {mmWave} new
  radio aided by sub-6 {GHz} {LTE} signaling,'' in
  \emph{https://arxiv.org/abs/1710.01879}, 2018.

\bibitem{va_impact_2017}
V.~Va, J.~Choi, and R.~W. Heath, ``The {Impact} of {Beamwidth} on {Temporal}
  {Channel} {Variation} in {Vehicular} {Channels} and {Its} {Implications},''
  \emph{{IEEE} Trans. Veh. Technol.}, vol.~66, no.~6, pp. 5014--5029, Jun.
  2017.

\bibitem{stabler_mimo_2009_ak}
O.~Stabler and R.~Hoppe, ``{MIMO} channel capacity computed with {3D} ray
  tracing model,'' in \emph{3rd {European} Conf. {Antennas} {Propagation}},
  Mar. 2009, pp. 2271--2275.

\bibitem{arikawa_simplified_2014}
S.~Arikawa and Y.~Karasawa, ``A {Simplified} {MIMO} {Channel} {Characteristics}
  {Evaluation} {Scheme} {Based} on {Ray} {Tracing} and {Its} {Application} to
  {Indoor} {Radio} {Systems},'' \emph{{IEEE} Antennas Wireless Propag. Lett.},
  vol.~13, pp. 1737--1740, 2014.

\bibitem{solomitckii_characterizing_2016}
D.~Solomitckii, Q.~C. Li, T.~Balercia, C.~R. C.~M. {da Silva}, S.~Talwar,
  S.~Andreev, and Y.~Koucheryavy, ``Characterizing the {Impact} of {Diffuse}
  {Scattering} in {Urban} {Millimeter}-{Wave} {Deployments},'' \emph{{IEEE}
  Wireless Commun. Lett.}, vol.~5, no.~4, pp. 432--435, Aug. 2016.

\bibitem{insite}
``Remcom {Wireless} {InSite},''
  \url{https://www.remcom.com/wireless-insite-em-propagation-software},
  accessed: 2018-01-20.

\bibitem{metis2015}
Mobile and wireless communications enablers for the Twenty-twenty
  information~society (METIS), ``Deliverable {D1.4} - {METIS} channel models
  (version 3),'' EU FP7 Project, Tech. Rep., 2015.

\bibitem{molisch_spatially_2016}
A.~F. Molisch, A.~Karttunen, S.~Hur, J.~Park, and J.~Zhang, ``Spatially
  consistent pathloss modeling for millimeter-wave channels in urban
  environments,'' in \emph{10th {European} Conf. {Antennas} {Propagation}
  ({EuCAP})}, Apr. 2016, pp. 1--5.

\bibitem{wang_extension_2016}
Y.~Wang, Z.~Shi, L.~Huang, Z.~Yu, and C.~Cao, ``An {Extension} of {Spatial}
  {Channel} {Model} with {Spatial} {Consistency},'' in \emph{Proc. {IEEE} 84th
  {Vehicular} {Technology} {Conference} ({VTC}-{Fall})}, Sep. 2016, pp. 1--5.

\bibitem{nyusim}
``{NYUSIM},'' \url{http://wireless.engineering.nyu.edu/nyusim}, accessed:
  2018-01-20.

\bibitem{Jaeckel14}
S.~Jaeckel, L.~Raschkowski, K.~Börner, and L.~Thiele, ``{QuaDRiGa}: A 3-d
  multi-cell channel model with time evolution for enabling virtual field
  trials,'' \emph{{IEEE} Trans. Antennas Propag.}, vol.~62, no.~6, pp.
  3242--3256, June 2014.

\bibitem{3gpp-R1-161622}
\BIBentryALTinterwordspacing
3GPP, ``{Spatial consistency modeling in drop based model},'' {3rd Generation
  Partnership Project (3GPP)}, Document for: Discussion and Decision R1-161622,
  2016. [Online]. Available:
  \url{https://portal.3gpp.org/ngppapp/CreateTdoc.aspx?mode=view&contributionId=691847}
\BIBentrySTDinterwordspacing

\bibitem{mckown_ray_1991}
J.~W. McKown and R.~L. Hamilton, ``Ray tracing as a design tool for radio
  networks,'' \emph{IEEE Network}, vol.~5, no.~6, pp. 27--30, Nov. 1991.

\bibitem{SUMO2012}
D.~Krajzewicz, J.~Erdmann, M.~Behrisch, and L.~Bieker, ``Recent development and
  applications of {SUMO - Simulation of Urban MObility},'' \emph{International
  Journal On Advances in Systems and Measurements}, vol.~5, no. 3\&4, pp.
  128--138, Dec. 2012.

\bibitem{Prelcic17}
N.~González-Prelcic, A.~Ali, V.~Va, and R.~W. Heath, ``Millimeter-wave
  communication with out-of-band information,'' \emph{{IEEE} Commun. Mag.},
  vol.~55, no.~12, pp. 140--146, Dec. 2017.

\bibitem{ali_millimeter_2017}
A.~Ali, N.~González-Prelcic, and R.~Heath, ``Millimeter {Wave}
  {Beam}-{Selection} {Using} {Out}-of-{Band} {Spatial} {Information},''
  \emph{{IEEE} Trans. Wireless Commun.}, vol.~17, no.~2, pp. 1038--1052, 2017.

\bibitem{va_inverse_2017}
V.~Va, J.~Choi, T.~Shimizu, G.~Bansal, and R.~W. Heath, ``Inverse {Multipath}
  {Fingerprinting} for {Millimeter} {Wave} {V}2i {Beam} {Alignment},''
  \emph{{IEEE} Trans. Veh. Technol.}, 2017, {IEEE} {Early} access.

\bibitem{kim_fast_2014}
J.~Kim and A.~F. Molisch, ``Fast millimeter-wave beam training with receive
  beamforming,'' \emph{Journal of Communications and Networks}, vol.~16, no.~5,
  pp. 512--522, Oct. 2014.

\bibitem{zhou_enhanced_2017}
P.~Zhou, X.~Fang, Y.~Fang, Y.~Long, R.~He, and X.~Han, ``Enhanced {Random}
  {Access} and {Beam} {Training} for {Millimeter} {Wave} {Wireless} {Local}
  {Networks} {With} {High} {User} {Density},'' \emph{{IEEE} Trans. Wireless
  Commun.}, vol.~16, no.~12, pp. 7760--7773, Dec. 2017.

\bibitem{kim_radio_1999}
S.-C. Kim, B.~J. Guarino, T.~M. Willis, V.~Erceg, S.~J. Fortune, R.~A.
  Valenzuela, L.~W. Thomas, J.~Ling, and J.~D. Moore, ``Radio propagation
  measurements and prediction using three-dimensional ray tracing in urban
  environments at 908 {MHz} and 1.9 {GHz},'' \emph{{IEEE} Trans. Veh.
  Technol.}, vol.~48, no.~3, pp. 931--946, May 1999.

\bibitem{5GMdata}
``{5GMdata},'' \url{https://github.com/aldebaro/5gmdata}, accessed: 2018-01-20.

\bibitem{dutta_frame_2017}
S.~Dutta, M.~Mezzavilla, R.~Ford, M.~Zhang, S.~Rangan, and M.~Zorzi, ``Frame
  {Structure} {Design} and {Analysis} for {Millimeter} {Wave} {Cellular}
  {Systems},'' \emph{{IEEE} Trans. Wireless Commun.}, vol.~16, no.~3, pp.
  1508--1522, Mar. 2017.

\bibitem{chollet2015keras}
F.~Chollet \emph{et~al.}, ``Keras,'' \url{https://github.com/keras-team/keras},
  2015.

\bibitem{plappert2016kerasrl}
M.~Plappert, ``{Keras-RL},''
  \url{https://github.com/matthiasplappert/keras-rl}, 2016.

\end{thebibliography}

\end{document}